# Resistance Switching Induced by Electric Field and Light Illumination in Device of FTO/CeO$_2$/Electrolyte/FTO

Xiaoqi Wang and Chuanbing Cai,

Physics Department, Shanghai University, Shanghai 200444, China

**Abstract:**

A heterojunction-like device consisting of FTO/CeO$_2$/electrolyte/FTO is established with distinct transport performance, where FTO denotes F-doped transparent conducting glass, and the electrolyte is LiI and I$_2$ in acetonitrile. The resistive switching behavior is observed, being induced through applying sufficient negative pulse as well as light illumination. The endurance measurements confirm that the write/erase periodic operation is reproducible and stable in the present device. Furthermore, the retention measurements demonstrate that the information can be stored temporarily for about 100 seconds. A possible mechanism regarding the formation of diiodide radical is proposed to give a reasonable explanation for the observed switching behavior.

Author's E-mail:
shawl.wang@gmail.com (Xiaoqi Wang)
cbcai@shu.edu.cn (Chuanbing Cai)

# 1. Introduction

Resistive switching (RS) behavior is an intriguing phenomenon in which the electrical resistance can be altered into a high resistance state (HRS) by a driving bias, and into a low resistance state (LRS) by another polarity bias for unipolar RS or by a reverse polarity bias for bipolar one [1, 2]. In the past decade, it has been investigated intensely, because of its potential application in nonvolatile memory devices and other electronic systems [3, 4]. Recently, great efforts are made with respect to a variety of heterostructures consisting of inorganic [5-7] or organic [8, 9] materials, where different mechanisms have been proposed to explain the RS phenomenon, such as charge ions trapped and detrapped, oxygen vacancies migration and deoxidization, and Schottky barrier [10].

Cerium dioxide ($CeO_2$) is an n-type semiconductor of an energy gap of 3.2 eV, with several advantages such as stable chemical properties and ease of processing, etc. In usual, it appears as an insulator as most metal electrodes such as Au and Pt have a deep Fermi level of 5.1~5.7 eV which is unfavorable for the carrier injection into the conduction band of $CeO_2$ with the electron affinity of about 3.5 eV [11]. In contrast, liquid electrolytes which are frequently used to study oxide semiconducting in the electrochemistry, exhibit the distinct electronic density of state with a Gaussian distribution in fluctuating energy levels [12]. This allows more opportunities for most oxide to capture (inject) carriers from (to) the conduction band (CB) or the valence band (VB). Of more interests, there are lots of oxides and electrolytes with various properties, allowing alternative novel combinations and low-cost architectures.

In the present work, we look into a special sandwich-type structure with $CeO_2$ and I-ionic electrolyte. As frequently used in dye-sensitized solar cells (DSSC), the electrolyte including the redox couple of $I^-$ and $I_3^-$ are purposely selected due to its stability of electronic transport and high reversibility [13]. The fabrication, performance and mechanism for the studied devices are addressed in details below.

## 2. Experimental

The polycrystalline $CeO_2$ film of around 200 nm in thickness was deposited on F-doped transparent conducting glass (FTO) by using pulsed laser deposition, in the oxygen atmosphere of 100 mTorr and at the substrate temperature of 500 ℃. The resultant $CeO_2$ film is filled with an electrolyte including 0.1 M LiI and 0.1 M $I_2$ in acetonitrile, and then sealed by a 25 μm-thick plastic spacer together with another FTO. The inset of Fig. 1 illustrates the fabricated device consisting of FTO/$CeO_2$/electrolyte/FTO with the $CeO_2$ area of 0.19 $cm^2$. Various voltage-current curves and endurance measurements are carried out by Keithley 2420 at the ambient temperature. The light source is provided by a commercial solar simulator (SAN-EI XES-151S) equipped with a 150W xenon lamp.

## 3. Results and discussion

### 3.1 Voltage-Current loops

Figure 1 displays a series of asymmetric non-linear *V-I* loop for the above electrochemical device. The bias is applied to the bottom electrode, and is initially swept through 0V→1V→ 0V→-1V→0V with a rate of 20 mV/s, seen in the inset of Fig.1. There is only a little hysteresis observed in the initial curve. After experiencing a negative bias, a pronounced hysteresis appears in the positive region. It is hard, however, to observe an obvious change in the negative bias region. Moreover, *V-I* loops show a good reproducibility during the 20 scanning, regardless of a little hysteresis for the initial scanning. This obviously evidences that the electric-induced resistive switching behaviors emerge in the present device.

### 3.2 Resistive switching induced by electric field or light illumination

To further investigate into the induced resistance switching behavior, the *V-I* curves in the positive bias region is recorded as well. Figure 2(a) shows the curves after a negative pulse of 5 s from -1 V to -2 V is applied. It is revealed that as the negative pulse is increased, the current is enhanced in forward sweeping (0→1 V), and reduced to a Schottky like behavior in the reverse (1→ 0 V). This suggests that the negative pulse drives the device into LRS while the positive bias recovers it to HRS. However, there is only a weak hysteresis observed during the negative bias region, together with a slight change in current produced after a sufficient positive pulse is applied. When the negative pulse is larger than -1.4 V, there is an additional drop can be observed during the scanning 0 V→0.1 V. This appears to be independent with the switching behavior, while its cause is not yet clear in reality.

Of most interests is that RS phenomenon can be induced not only by the electric field, but also by the light illumination. Fig. 2(b) illustrates the switching behavior induced by sun light illumination with various durations. As the duration is prolonged, RS phenomenon becomes significant. For comparison, the switching behavior due to applying -1 V pulses with various durations is indicated in the Fig. 2(c). These imply that the RS phenomenon can be induced and recollected by the external condition, i.e. the electric pulse and light illumination, but being independent on ion diffusion due to the electrostatic force.

*3.3 Operation performances*

It is more interesting that whether or not the present electrochemical device can perform well in practical operations. To ascertain the endurance, we apply a negative pulse as "write", i.e. -1 V, to switch it into LRS, and a positive one as "erase", i.e. +1 V, to recover it into HRS. After each write (erase) operation, the low (high) resistance states are read out (at +0.3 V). The time duration of each operation is set as 0.1s. Fig. 3(a) shows the results of the 5000-cycle endurance measurement, demonstrating the stability and reproducibility for repeated operation, regardless of a little hysteresis observed in the first 100 cycles. Note that the low resistance is about 5 KΩ and the high resistance is above 80 KΩ, as shown in the inset of Fig. 3(a). Correspondingly, the ratio of HRS and LRS is in excess of 10 times, suggesting the potential application of such an electrochemical device.

To ascertain the retention properties, the present device cell is read out at the 0.3 V after a "write" at -1 V or an "erase" at 1 V. As illustrated in the Fig. 3(b), two current states can be distinguished after applying "write" and "erase" by 10 seconds.

The higher current decays quickly and converges with the lower one for delays larger than 100 s after the pulses, implying the device can temporarily recollect the information for about 100s. Although the retention time present is rather short, the potential application such as a novel memory cell based on electrochemical device can be expected definitely.

*3.4 Potential switching mechanisms*

Till now, we have not yet discussed the mechanisms for the observed switching behavior. According to prior literatures [10], a few models regarding the barrier at interface of heterojunction are applicable to explain RS phenomenon. However, for the present device with a peculiar electrochemical structure, it seems more complicate, requiring additional experiments to manifest the critical role and effect of interfaces among the FTO/$CeO_2$, $CeO_2$/electrolyte, and electrolyte/FTO on the switching performance. Firstly, a simple structure without $CeO_2$, FTO/electrolyte/FTO is prepared, and further experiments clarify that there is no switching phenomena no matter with or without a sufficient bias applied. Secondly, an electrolyte-free architecture, FTO/$CeO_2$/ITO is prepared. And no RS behavior is observed as well. Thus both electrolyte/FTO and FTO/$CeO_2$ interfaces can be excluded.

Regarding the barrier model, it is assumed that the junction barrier is gradually transformed from the diffusive to the tunneling type during the reverse sweeping (0→-1V). And it is transformed back to a diffusion one again during the forward sweeping (-1V→0V), leading to the current higher than that in the reverse sweeping [7]. This result, however, is contrary to our experimental observation which indicates that the current in reverse sweeping is a little higher than that in the forward, also being hard to explain the switching induced after illumination. Therefore, it is reasonable to rule out the effect of the emerged barrier.

*3.4 Model regarding the formation of diiodine radicals*

It is worth noting that the redox reaction at the interface of $CeO_2$/electrolyte, involves the constituent process of two electrons transferred from $I^-$ to $I_3^-$ through a intermediate species, i.e. diiodide radicals $I_2^{-*}$, as shown in equation (1)-(3). The intermediate species may appear through a charge transfer of low electron-density [12] or through a laser illumination [14-16]. Because that the switching in the present device can be induced by electric pulse and illumination, one may propose that RS behavior is subject to the existing diiodide radical produced induced by sufficient negative biases and by illumination. In reality, the above proposal can explain the experimental data in qualitatively, although further experiments are required to understand the mechanism.

$$2I^- - e^- \rightleftharpoons I_2^{-*} \quad (1)$$

$$I^- + I_2^{-*} - e \rightleftharpoons I_3^- \quad (2)$$

$$I_3^- + e^- \rightleftharpoons I_2^{-*} + I^- \quad (3)$$

In the above-mentioned experiments, only the case of *V-I* sweeping is discussed. Figure 4 illustrates the energy level diagram for the interface of $CeO_2$/electrolyte in the case of electric measurements, where the energy level of FTO is omitted for simplification. As the initial sweeping from 0 V, electrons are transferred from $I^-$ to the CB or the surface states (SS) of $CeO_2$, performing a Schottky-type current shown in the Fig.1. Although the $I_2^{-*}$ are produced, they will quickly inject electrons to $CeO_2$ and no radicals are stored, as shown in Fig. 4(a).

In the case of reversely sweeping, however, the electrons can be injected from $CeO_2$ to electrolyte and then reduce $I_3^-$ into $I^-$ and $I_2^{-*}$, following the equation (3). The formed $I_2^{-*}$ will be temporally stored because it is hard to transfer the electrons due to the mismatch of energy levels between $CeO_2$ and $I_2^{-*}$, as shown in Fig. 4(b). Furthermore, as the bias forwardly sweeping again, the stored $I_2^{-*}$ radicals transfer the electrons to the $CeO_2$ until its concentration is exhausted, leading to the current increase up to a peak and then decrease as the Schottky type, which are corresponding to LRS and HRS, respectively.

## 4. Conclusions

In summary, it is demonstrated that the resistive switching behavior is reproducible in an electrochemical heterojunction consisting of FTO/$CeO_2$/electrolyte/FTO. Electrical transport measurements show that the switching behavior is subject to the applied negative pulse, as well as the light illumination. The endurance measurements confirm the stability in repeated operation, and the retention measurements demonstrate the properties of information storage, suggesting the potential applications as a novel electrochemical device, although its retention time is still short. After ruling out several possible mechanisms, it is believed that the formation and storage of diiodide radicals are responsible for the observed switching behavior, which appears tailorable by applying sufficient negative bias or by light illumination,

## Acknowledgement:

This work is partly sponsored by the Innovation Funds for Ph.D. Graduates of Shanghai University (2011), China.

**FIGURE CAPTIONS:**

**Fig. 1**. Twenty scanning of *V-I* loop for the FTO/CeO$_2$/electrolyte/FTO structure. Inset is the illustration of the present junction-like device.

**Fig. 2** *V-I* scanning of the studied device after different treatments. (a) applying various negative pulses for 5s; (b) illumination with a simulated sunlight of AM 1.5; (c) applying -1 V of various durations.

**Fig. 3.** (a) 5,000-cycle Endurance measurement for the "write-read-erase–read" operation. The first 100-cycle is enlarged in the inset. (b) Retention measurement with the temporal width of read pulses at 0.1 s. The applied voltage is fixed to zero between each read pulses.

**Fig. 4.** Schematic diagrams for the energy levels of CeO2/electrolyte. (a) initially scanning; (b) reversely scanning; (c) forwardly scanning. The multi-lines denote the surface states. The dotted lines represent the Fermi level in CeO2 and the redox level in electrolyte, respectively. The arrows imply the direction of electron transfer.

Fig. 1

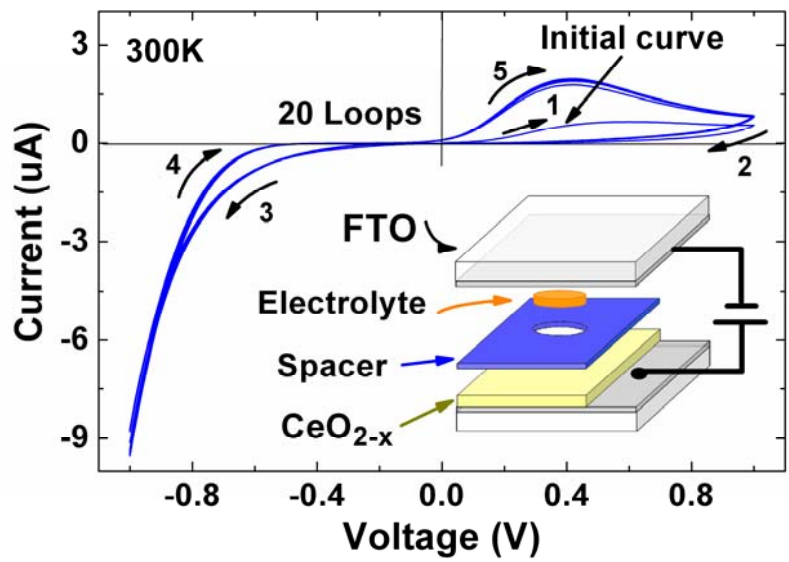

Fig. 2(a), (b) & (c)

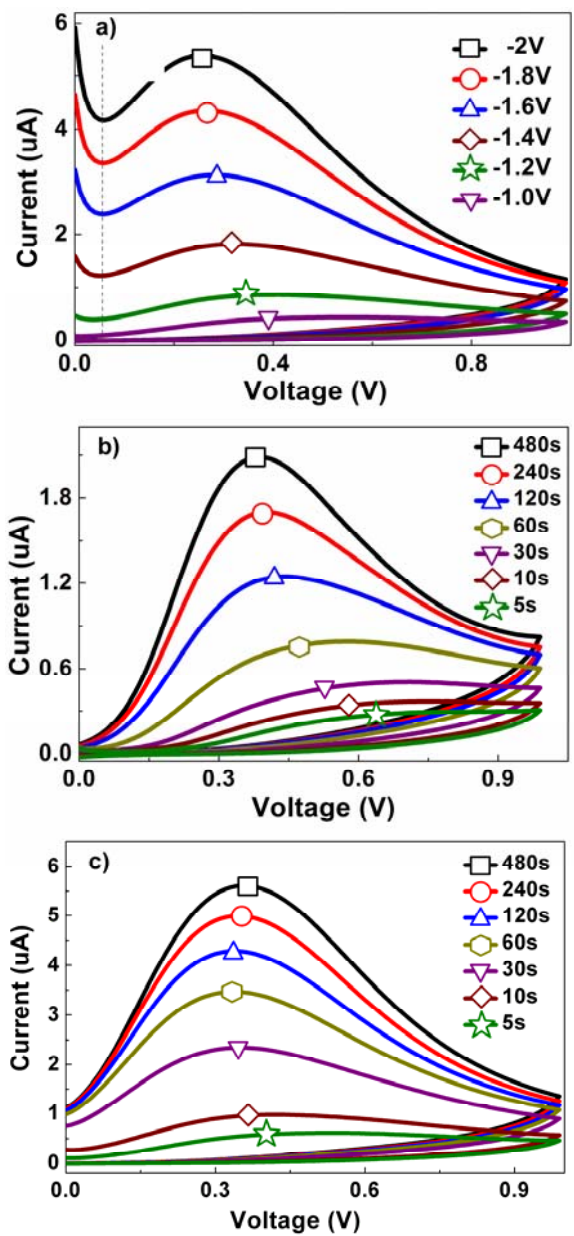

Fig 3(a) & (b)

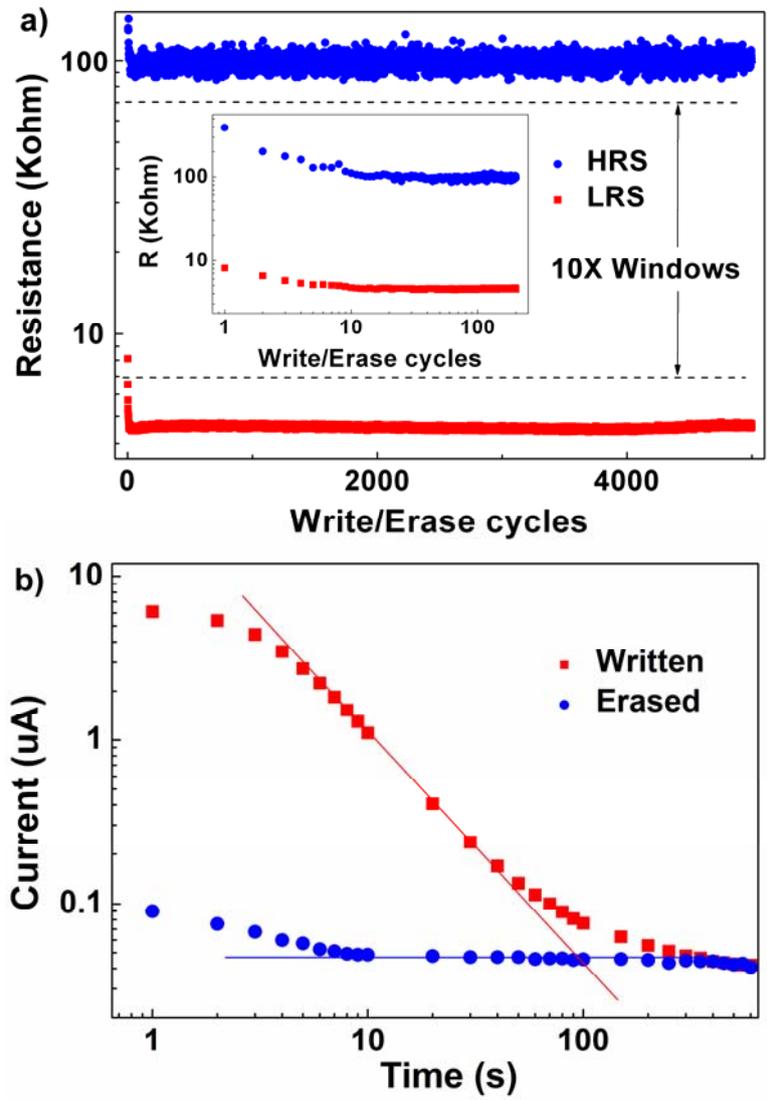

Fig. 4

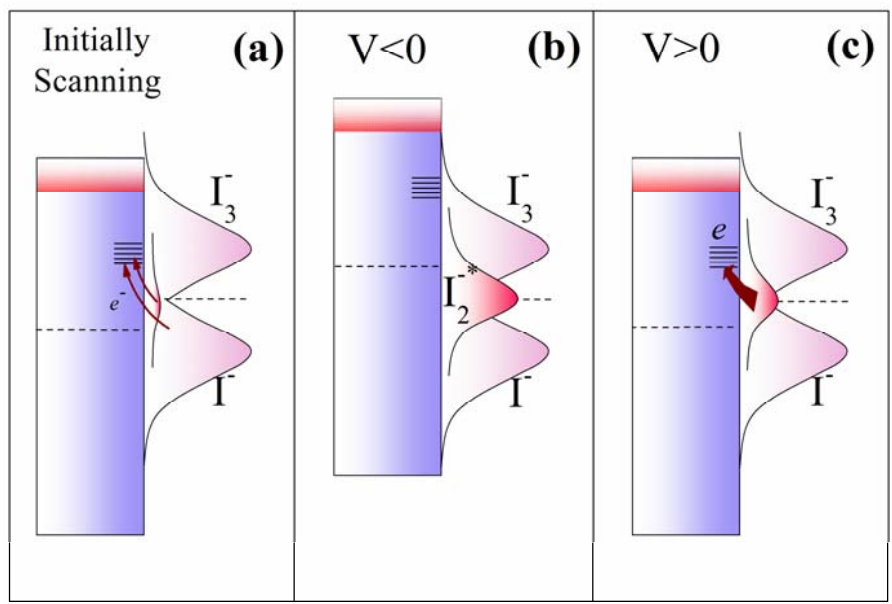